\documentclass[floatfix,twocolumn,superscriptaddress]{revtex4-2}
\usepackage{xcolor}
\usepackage{mathrsfs}
\usepackage{graphicx}
\usepackage[english]{babel}
\usepackage{amsmath}
\usepackage{amssymb}
\usepackage{booktabs}
\usepackage{array}
\usepackage{tabularx}
\usepackage{makecell}
\usepackage{relsize}
\usepackage{CJK}
\usepackage{dsfont}
\usepackage{bm}
\usepackage{xcolor}

\newcommand{\me}{\mathrm{e}}

\allowdisplaybreaks
\begin{document}

\title{Curvature-driven shifts of the Potts transition on spherical Fibonacci graphs: a graph-convolutional transfer-learning study}

\author{Zheng Zhou }
\affiliation{School of Physics, Southeast University, Jiulonghu Campus, Nanjing 211189, China}

\author{Xu-Yang Hou}
\affiliation{School of Physics, Southeast University, Jiulonghu Campus, Nanjing 211189, China}

\author{Hao Guo}
\email{guohao.ph@seu.edu.cn}
\affiliation{School of Physics, Southeast University, Jiulonghu Campus, Nanjing 211189, China}
\affiliation{Hefei National Laboratory, Hefei 230088, China}

\begin{abstract}

We investigate the ferromagnetic $q$-state Potts model on spherical Fibonacci graphs. These graphs are constructed by embedding quasi-uniform sites on a sphere and defining interactions via a chord-distance cutoff chosen to yield a network approximating four-neighbor connectivity. By combining Swendsen-Wang cluster Monte Carlo simulations with graph convolutional networks (GCNs), which operate directly on the adjacency structure and node spins, we develop a unified phase-classification framework applicable to both regular planar lattices and curved, irregular spherical graphs. Benchmarks on planar lattices demonstrate an efficient transfer strategy: after a fixed binarization of Potts spins into an effective Ising variable, a single GCN pretrained on the Ising model can localize the transition region for different $q$ values without retraining. Applying this strategy to spherical graphs, we find that curvature- and defect-induced connectivity irregularities produce only modest shifts in the inferred transition temperatures relative to planar baselines. Further analysis shows that the curvature-induced shift of the critical temperature is most pronounced at small $q$ and diminishes rapidly as $q$ increases; this trend is consistent with the physical picture that, in two dimensions, the Potts model undergoes a transition from a continuous phase transition to a weakly first-order one for $q>4$, accompanied by a pronounced reduction of the correlation length.

\end{abstract}

\maketitle

\section{Introduction}
In recent years, machine learning has developed into a powerful tool for identifying phase transitions directly from raw configurations, serving as an important complement to traditional thermodynamic approaches. Early pioneering studies demonstrated that neural networks can automatically detect qualitative changes in configurations near phase transitions without explicit knowledge of order parameters \cite{MLS17,MLNP17,MLPRX17b,MLPRL18a,PhysRevX.7.031038}. Building on these insights, subsequent works systematically extended this idea to a variety of classical and quantum models, establishing the applicability and stability of machine-learning methods across diverse physical systems \cite{LI2018312,PhysRevE.98.022138,PhysRevB.94.195105,PhysRevB.99.075418}. These approaches were further generalized to systems in different spatial dimensions, demonstrating strong robustness and generalization with respect to dimensionality \cite{PhysRevE.102.053306,PhysRevB.100.045129,PhysRevResearch.2.023266}. In addition, a series of studies investigated the robustness of machine-learning models against noise and finite-size effects, showing that reliable phase identification remains possible even under non-ideal data conditions \cite{PhysRevE.96.022140,PhysRevE.97.032119,JMLR:v18:17-527}. Beyond qualitative classification, further research explored the quantitative extraction of critical points and effective critical exponents using machine-learning techniques, thereby elevating these methods into quantitative tools for analyzing phase transition behavior \cite{WOS:000400149400001,WOS:000400476900008,WOS:000598326500001,WOS:000689499400001,WOS:000853261600014}.

Most early successes employed convolutional neural networks (CNNs), which assume image-like data on regular grids ~\cite{CNNPRB18,CNNPRE19, WOS:000400149400001}. However, for curved surfaces and irregular lattices, the data are not arranged on a translationally invariant grid; instead, the notion of ``nearest neighbors" is encoded by an interaction graph. This mismatch motivates the use of graph based learning models. Graph convolutional networks operate directly on a graph adjacency structure and can thus treat regular and irregular lattices within a unified framework ~\cite{PhysRevResearch.4.023005,GCN20,Zhou_2025}.

Phase transitions on curved or irregular lattices also provide a setting to probe how geometry and connectivity shape collective phenomena beyond the standard Euclidean paradigm. In many contexts, such as spin degrees of freedom constrained on curved manifolds, or interaction networks without a regular embedding, the system is not a translationally invariant Bravais lattice, and ``nearest neighbors" are not uniquely defined by a single bond length ~\cite{Kleman_1989,Janke_2002,giomi2009unordinary}. A natural approach is to treat the system as a statistical model on a graph, where geometric information enters primarily through connectivity.

Among discrete spin models, the ferromagnetic $q$-state Potts model is a canonical extension of the Ising model, exhibiting rich critical behavior on two-dimensional lattices. On an infinite planar square lattice, the critical temperature is exactly known, $T_c^{2\mathrm{D}}(q)/J=1/\ln(1+\sqrt{q})$, with a continuous transition for $q\le 4$ and a first-order transition for $q > 4$ ~\cite{Wu_1982}. These exact planar results provide a well-established benchmark, making the Potts model an ideal testbed for assessing how curvature and irregular coordination modify critical properties on a spherical graph. The precise knowledge of $T_c^{2\mathrm{D}}(q)$ offers a clear reference against which shifts induced by non-Euclidean geometry can be quantified.

A practical challenge is to discretize the sphere uniformly enough for meaningful comparison with planar lattices while keeping the construction simple for arbitrary $N$. We adopt the spherical Fibonacci lattice ~\cite{WOS:000273034500003,Hannay_2004,WOS:000363671200030}, which provides quasi-uniform coverage through a closed form construction. Any local interaction graph built on such a point set inevitably contains coordination defects. To emulate a ``square-lattice-like" local environment, we define the interaction graph by a chord distance cutoff $r_c$, chosen to maximize the number of four-neighbor nodes. This yields predominantly four-neighbor spherical graphs with a controlled fraction of three and five neighbor sites. This construction extends our recent spherical Fibonacci lattice Ising study ~\cite{Zhou_2025} and provides a consistent baseline for generalizing to $q > 2$.

To sample equilibrium Potts configurations efficiently across temperatures, we use the Swendsen-Wang cluster Monte Carlo algorithm ~\cite{Edwards_Sokal_1988,Galanis_Stefankovic_Vigoda_2019,Barbu_Zhu_2005,Blanca_Gheissari_2023,Ivaneyko_2005}. This method updates collective Fortuin-Kasteleyn (FK) clusters, mitigating critical slowing down near the transition. Beyond improving sampling efficiency, the FK representation offers an intuitive visualization of connectivity and domain formation on irregular graphs.
For phase classification, we employ GCNs: for a fixed interaction graph, each configuration is represented by node spins and the adjacency matrix, and the GCN outputs confidence scores for the ordered and disordered phases. We adopt an ``extreme-temperature" supervised training protocol, where configurations from very low and very high temperatures are labeled as ordered and disordered, respectively. The trained network is then evaluated across the full temperature range, and the transition region is identified where the two output confidences become comparable. To avoid training a separate network for every $q$, we introduce a simple binarization rule that maps the $q$-state Potts spins into an effective Ising variable. This allows a GCN pretrained on the Ising model to be applied directly to binarized Potts configurations for different $q$, preserving the large-scale domain geometry relevant for phase classification. We first validate this transfer strategy on the planar square lattice using the exact $T_c^{2\mathrm{D}}(q)$ as a benchmark, and then apply the same framework to spherical Potts models on Fibonacci graphs.

The main goals of this work are threefold:
(i) to construct and characterize nearly four-neighbor spherical Fibonacci graphs suitable for Potts simulations;
(ii) to develop a unified GCN based phase classification framework applicable to both planar and spherical lattices, including an efficient transfer strategy across $q$ based on binarization;
(iii) to quantify how curvature and local connectivity affect Potts critical temperatures on the sphere relative to planar baselines.
By combining controlled graph construction, Swendsen-Wang sampling, and graph based learning, we systematically compare $T_c(q)$ across geometry (plane vs. sphere), spin multiplicity, and local connectivity (tuned via $r_c$).

\section{Spherical Fibonacci lattice}\label{II}

\begin{figure}[ht]
\centering
\includegraphics[width=1.5in]{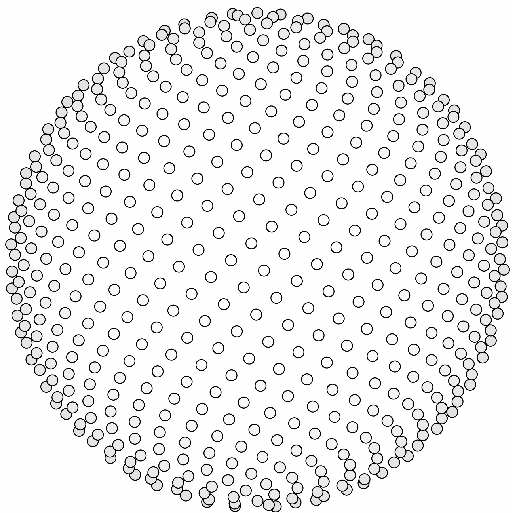}
\hspace{0.2in}
\includegraphics[width=1.5in]{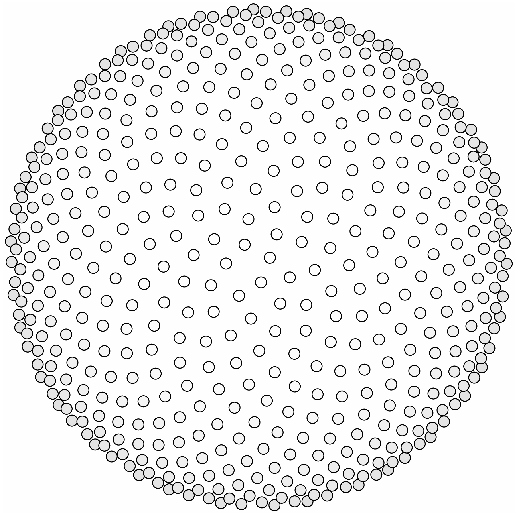}
\caption{Spherical Fibonacci sites for $N=1000$ ($R=10$), shown in orthographic front (left) and top (right) views.}
\label{fig:lattice}
\end{figure}

We discretize a spherical surface of radius $R$ by placing $N$ sites using the spherical Fibonacci construction ~\cite{Hannay_2004}, which provides a quasi uniform point set. The Cartesian coordinates of the $i$ th site ($i=1,2,\dots,N$) are
\begin{align}\label{eq:fib_sites}
x_i &= \sqrt{R^2-z_i^2}\,\cos\theta_i,\qquad
y_i = \sqrt{R^2-z_i^2}\,\sin\theta_i,\notag\\
z_i &= R\left(\frac{2i-1}{N}-1\right),\qquad
\theta_i = 2\pi i\,\varphi,
\end{align}
where $\varphi=(\sqrt{5}-1)/2$ is the inverse golden ratio.
Fig.~\ref{fig:lattice} shows the resulting Fibonacci sites for $N=1000$ in two orthographic views (front and top).

To define short range interactions on this irregular point set, we construct an interaction graph by introducing a cutoff radius $r_c$: two sites $i$ and $j$ are connected if their chord distance
\begin{equation}
d_{ij}=|\boldsymbol r_i-\boldsymbol r_j|
\end{equation}
satisfies $d_{ij} < r_c$.

\begin{figure}[ht]
\centering
\includegraphics[width=1.5in]{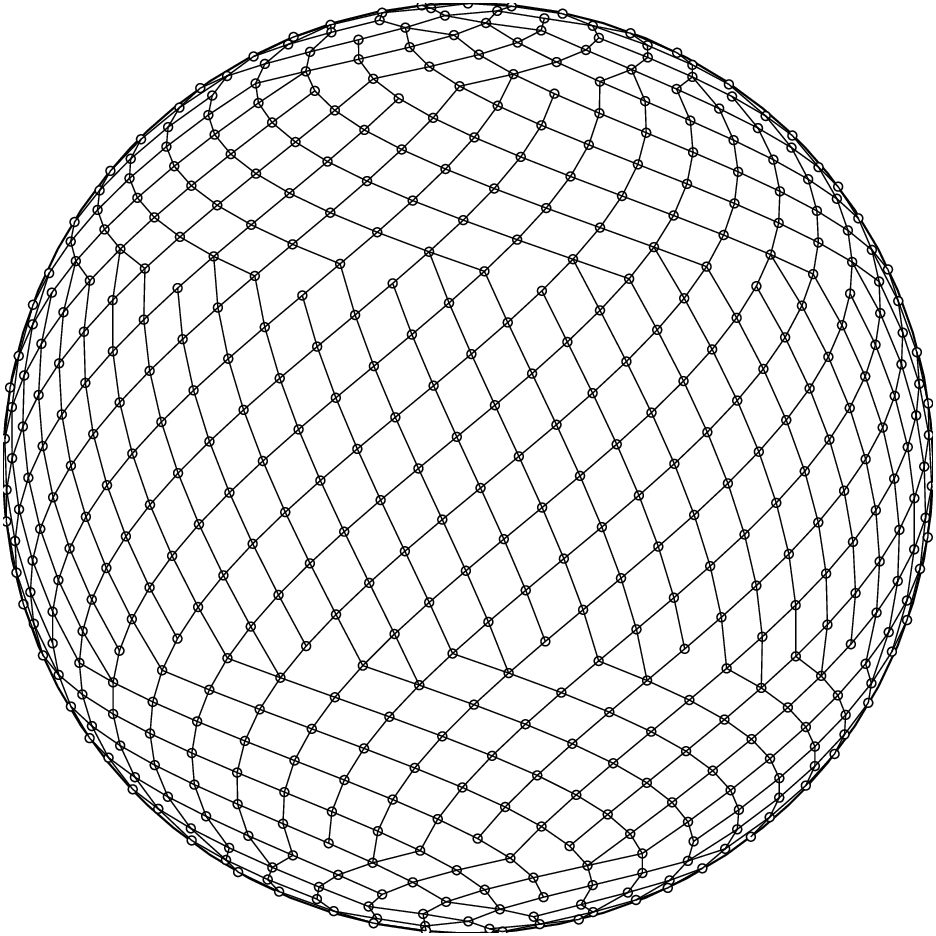}
\hspace{0.2in}
\includegraphics[width=1.5in]{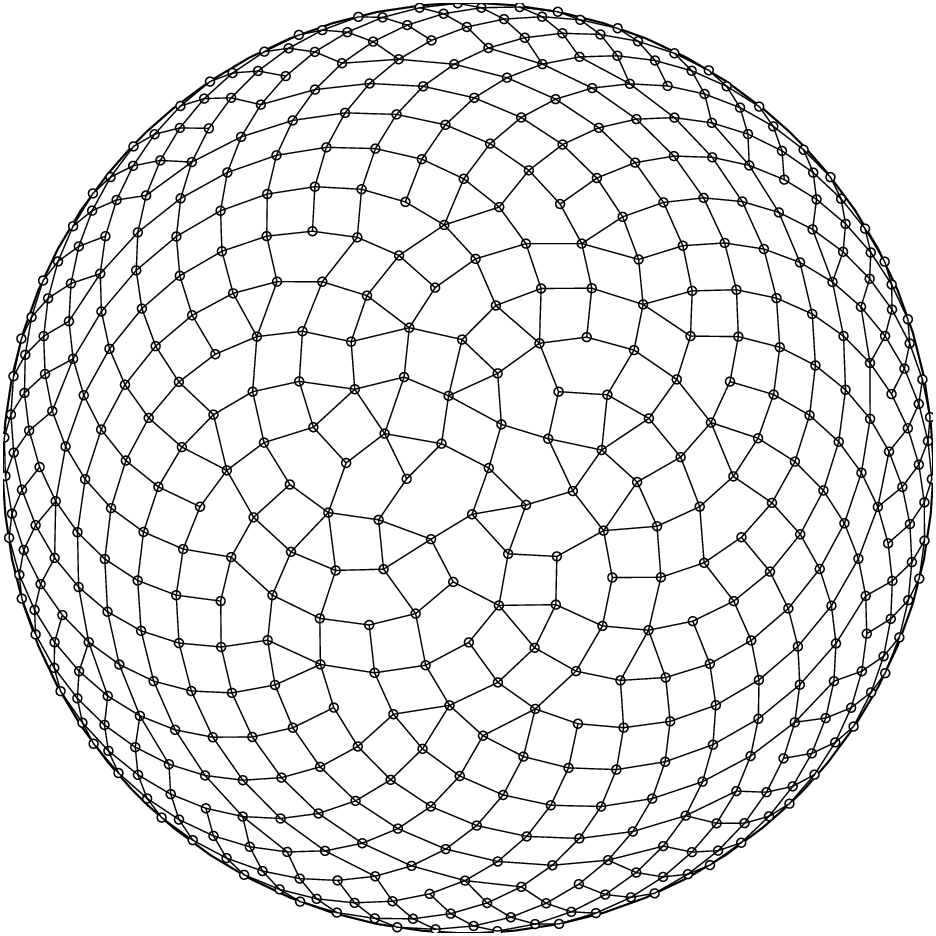}
\caption{Interaction graph constructed from the Fibonacci sites in Fig.~\ref{fig:lattice} using $r_c/R=0.1298$:
edges connect pairs with $d_{ij}<r_c$ (orthographic front/top views).}
\label{fig:connect}
\end{figure}

In numerical simulations, once the sphere radius $R$ and the total number of sites $N$ are fixed, we typically choose an appropriate cutoff radius ($r_c$) (or equivalently the dimensionless ratio $r_c/R$) such that the resulting interaction graph is approximately four-neighbor. As a concrete example, for $N=1000$ and $R=10$, choosing $r_c/R=0.1298$ (i.e., $r_c=1.298$) yields a system in which the vast majority of sites are four-neighbor. The coordination statistics are $(N_3,N_4,N_5)=(76,850,74)$, corresponding to an average coordination number
$\bar z=(3N_3+4N_4+5N_5)/N\simeq 3.998$. Under this construction, the lattice contains only a small fraction of three- and five-neighbor sites as unavoidable coordination defects, while remaining a close approximation to a four-neighbor lattice. Fig.~\ref{fig:connect} shows the resulting connectivity from the same two views as in Fig.~{\ref{fig:lattice}}.


\section{The $q$-state Potts model on spherical Fibonacci lattice}\label{III}
\subsection{Spin configurations}

The 
$q$-state Potts model is a classical spin model in which each site $i$ carries a discrete spin variable $\sigma_i\in\{0,1,\dots,q-1\}$. The Hamiltonian is defined as
\begin{equation}
  H = -J \sum_{\langle i,j\rangle} \delta_{\sigma_i,\sigma_j},
  \label{eq:H_potts_cn}
\end{equation}
where $\langle i,j\rangle$ runs over all interacting pairs with $d_{ij} < r_c$, $J > 0$ is the coupling constant, and $\delta$ is the Kronecker delta. The case $q=2$ reduces to the Ising model. We measure the temperature $T$ in units of $J/k_B$, and the corresponding Boltzmann weight is given by
\begin{equation}
  P(\{\sigma_i\}) \propto
  \exp\!\left(\frac{J}{T}\sum_{\langle i,j\rangle}
  \delta_{\sigma_i,\sigma_j}\right).
\end{equation}
For convenience, natural units are adopted in the following, with the Boltzmann constant set to $k_B=1$. On the infinite planar square lattice, the critical temperature of the $q$-state Potts model is exactly $T_c^{2\mathrm{D}}(q)/J=1/\ln(1+\sqrt{q})$ ~\cite{Wu_1982}, with the transition being second order for $q\le 4$ and first-order for $q > 4$. In this work, we investigate the phase transition properties of the $q$-state Potts model on the spherical Fibonacci lattice.

We generate equilibrium configurations using the Swendsen-Wang cluster Monte Carlo algorithm ~\cite{Edwards_Sokal_1988,Galanis_Stefankovic_Vigoda_2019,Barbu_Zhu_2005,Blanca_Gheissari_2023,Ivaneyko_2005}. Within the Fortuin-Kasteleyn (FK) representation, each nearest-neighbor edge $\langle i,j\rangle$ is activated with probability $p_{\mathrm{add}} = 1 - e^{-J/T}$ if $\sigma_i=\sigma_j$, and remains inactive if $\sigma_i\neq\sigma_j$. These activated edges connect like-spin sites into FK clusters. Each cluster is then collectively reassigned a new spin value, drawn uniformly from $\{0,\dots,q-1\}$. This cluster flip preserves detailed balance and significantly suppresses critical slowing down near $T_c$.

In numerical simulations, for each set of parameters $(R,N,r_c)$ and $q$, Swendsen–Wang cluster simulations are performed over a common temperature range $T/J\in[0.01,4.0]$, with a step of $\Delta T/J=0.01$.
At each temperature, we start from a random initial configuration, perform $N_{\mathrm{eq}}=2000$ equilibration sweeps, and then record $N_{\mathrm{samp}}=100$ configurations separated by an interval of 40 sweeps. The saved configurations are treated as approximately uncorrelated samples.

\begin{figure}[t]
  \centering
\includegraphics[width=1.6in]{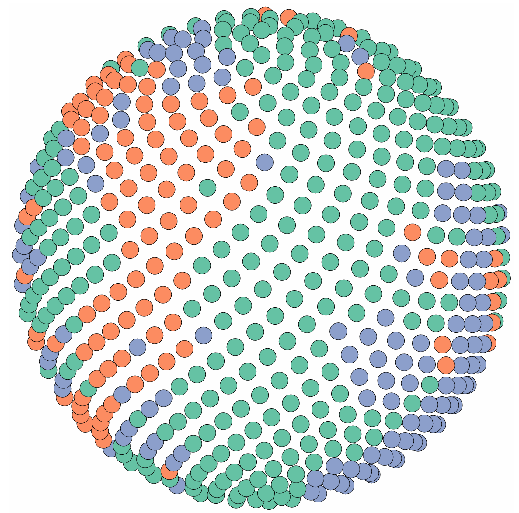}
\hspace{0.1in}
\includegraphics[width=1.6in]{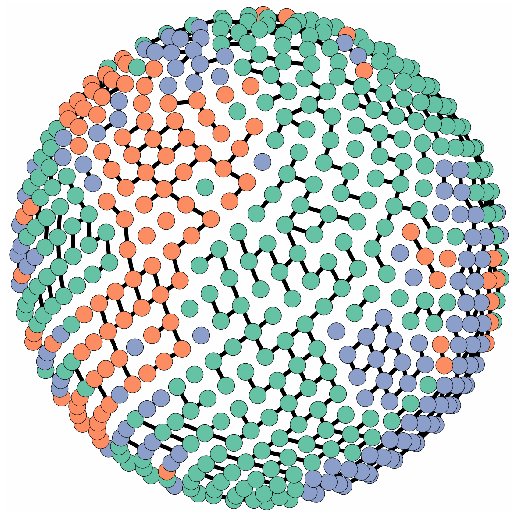}
  \caption{Bond activation in the Swendsen--Wang algorithm ($N=1000$, $R=10$, and $r_c/R=0.1298$).
  Left: a representative spin configuration at $T/J=1.0$ for $q=3$.
  Right: the same configuration overlaid with the activated FK bonds, which
  connect equal-spin neighbors into FK clusters.}
  \label{fig:sphere_plane_bonds_cn}
\end{figure}

Both the equilibrium spin configurations and the corresponding activated FK bonds are recorded; the latter are used to visualize cluster connectivity (Fig.~\ref{fig:sphere_plane_bonds_cn}). These same configurations serve as input for the subsequent machine-learning analysis. Fig.~\ref{fig:sphere_plane_bonds_cn} illustrates the FK cluster geometry for $q=3$ at $T/J=1.0$, obtained on a spherical Fibonacci lattice with parameters $N=1000$, $R=10$, and $r_c/R=0.1298$: the left panel displays the spin configuration, while the right panel overlays the activated FK bonds from one Swendsen-Wang update. Because bonds are activated exclusively between spins of the same value, the bond network directly outlines the FK clusters and provides a visualization of the correlation structure.

\subsection{Planar benchmarks and binarization-based GCN estimator}

Before applying GCNs to the spherical Potts model, we first validate the method on the two-dimensional square lattice. A square lattice can be represented as a graph where lattice sites are nodes and nearest neighbor couplings are edges. We use the same GCN architecture as in our spherical Ising analysis ~\cite{Zhou_2025}: the inputs are the spin configuration and the adjacency matrix, and the outputs are two components $(p_\text{o},p_\text{d})$ representing classification confidence for the ordered and disordered phases. Compared to convolutional neural networks on planar images, GCNs operate directly on graph structures and treat regular and irregular lattices within a unified framework ~\cite{PhysRevResearch.4.023005}.

\textit{Direct training on planar $q=3$.}
We first train a GCN directly on the 2D $q=3$ Potts model on a square lattice of linear size $L=128$. Using the Swendsen-Wang algorithm, we generate equilibrium configurations at extremely low temperatures $T/J\in[0.01,0.05]$ (labeled as ordered with $(p_\text{o},p_\text{d})=(1,0)$) and extremely high temperatures $T/J\in[100,104]$ (labeled as disordered with $(p_\text{o},p_\text{d})=(0,1)$) ~\cite{MLS17}. As node features we use the raw Potts spin values $\sigma_i\in\{0,1,2\}$ encoded as a scalar per node. After training converges, we freeze the network and perform inference across $T/J\in[0.01,4.0]$ to obtain $p_\text{o}(T)$ and $p_\text{d}(T)$.

\begin{figure}[t]
  \centering
  \includegraphics[width=1\linewidth]{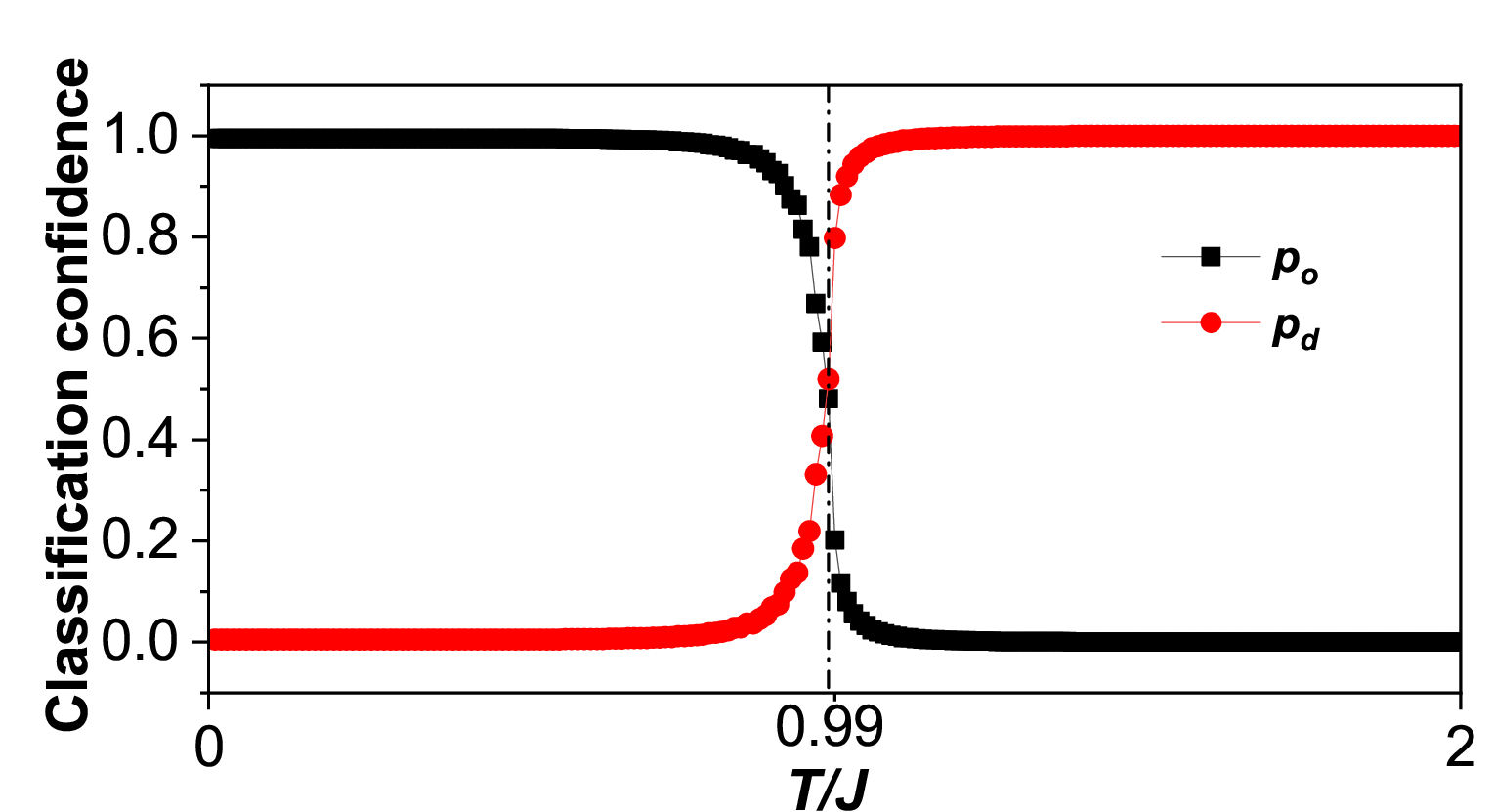}
\caption{GCN confidence for the 2D $q=3$ Potts model ($L=128$). }
  \label{fig:planar_q3_confidence_raw_cn}
\end{figure}

Fig.~\ref{fig:planar_q3_confidence_raw_cn} displays the confidence curves obtained for $q=3$. A monotonic decrease in $p_\text{o}(T)$ and a concomitant increase in $p_\text{d}(T)$ are observed with increasing $T$. At the resolution of our simulation ($\Delta T = 0.01J$), the curves intersect at $T/J \simeq 0.99$, closely matching the exact critical value of $T_c^{2\mathrm{D}}(3)/J\approx 0.9950$. This benchmark validates that the GCN, trained exclusively on extreme-temperature configurations, can correctly identify the transition thereby capturing the underlying change in domain structure.

\textit{Binarization and transfer learning.}
Training a separate network for each $q$ is computationally inefficient. Recent CNN studies suggest that after an appropriate binarization of Potts spins into two groups, a network trained on the Ising model can transfer to the $q$-state Potts model without losing critical information ~\cite{Fukushima_Sakai_2021}. Motivated by this, we adopt a binarization strategy in the GCN setting. For a given $q$, we partition the $q$ spin states into two sets and define an effective Ising variable $s_i=\pm1$ as
\begin{equation}
s_i=
\begin{cases}
-1, & \sigma_i \in \{0,1,\dots,\lfloor (q-1)/2\rfloor\},\\
+1, & \sigma_i \in \{\lfloor (q-1)/2\rfloor+1,\dots,q-1\}.
\end{cases}
\label{eq:binarize_rule}
\end{equation}
For example, for $q=3$ we choose $s_i=-1$ for $\sigma_i=0$ and $s_i=+1$ for $\sigma_i=1,2$. This mapping coarsens the $q$ valued Potts field into a binary one while leaving the interaction graph unchanged. At the cluster level, it approximately preserves the large scale geometry of domains associated with the chosen partition, specifically the interfaces separating the two groups, while discarding distinctions between Potts states within the same group.

\begin{figure}[htbp]
\centering
\includegraphics[width=3.5in]{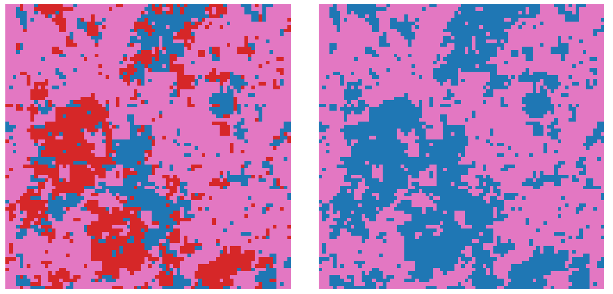}\\
\caption{Planar Potts configurations (q=3) and their binarized versions at $T/J=1.0$.
Left: Potts spins $\{\sigma_i\}$; right: binarized Ising spins $\{s_i=\pm1\}$ (see Eq.~\ref{eq:binarize_rule}).}

\label{2D_binarized_q3}
\end{figure}

\begin{figure}[t]
  \centering
  \includegraphics[width=1\linewidth]{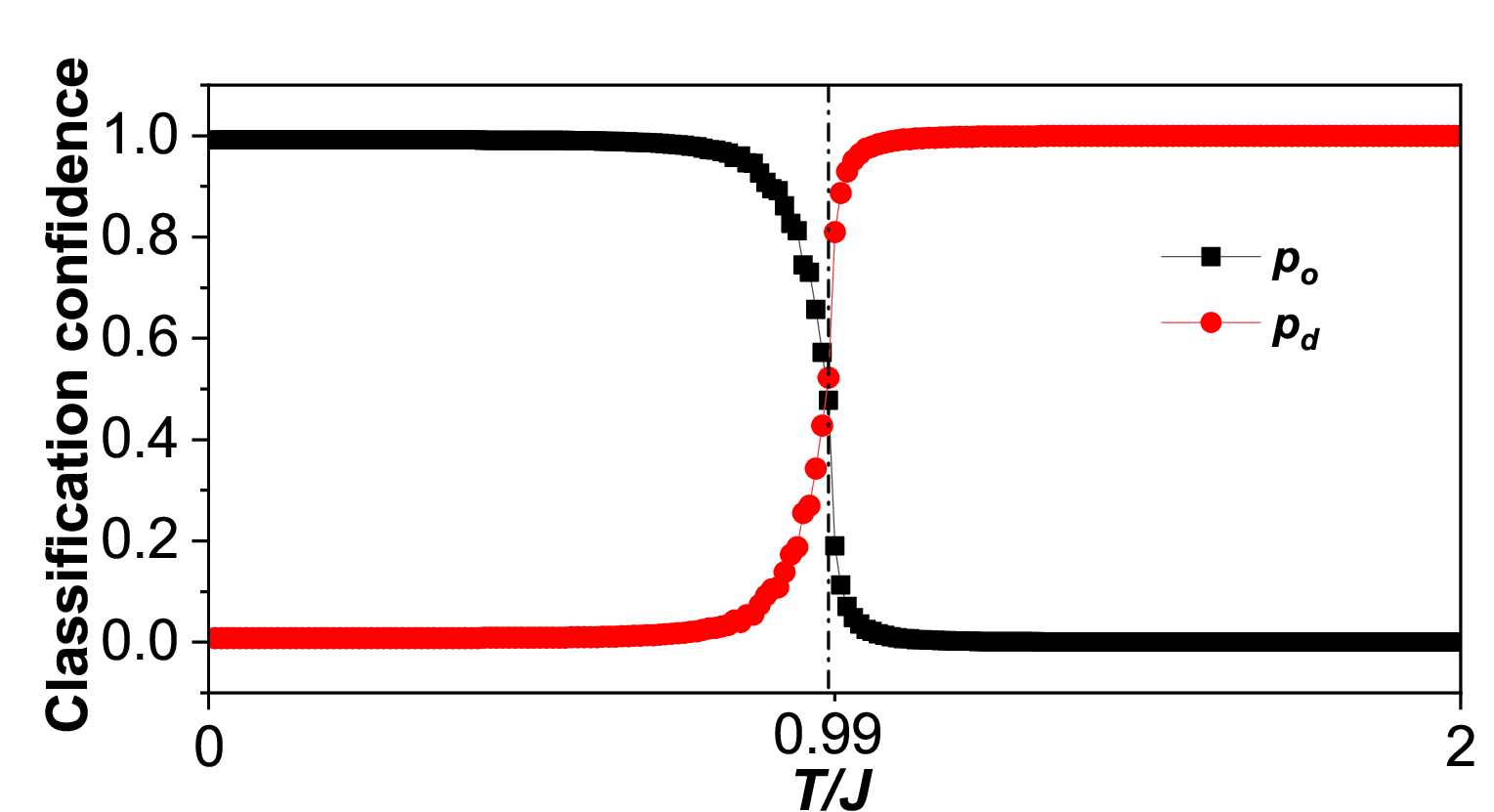}
  \caption{GCN confidence curves for the 2D $q=3$ Potts model ($L=128$) after binarization.}
  \label{fig:planar_q3_confidence_binary_cn}
\end{figure}

Keeping the network architecture fixed, we recompute the confidence curves for the $q=3$ dataset using the binarized $s_i$ as inputs. As shown in Fig.~\ref{fig:planar_q3_confidence_binary_cn}, the crossing remains at $T_c/J\approx 0.99$, indicating that binarization does not destroy the critical information carried by Potts cluster geometry. For a visual demonstration, Fig.~\ref{2D_binarized_q3} compares original Potts configurations and their binarized counterparts at $T/J=1.0$ for $q=3$: large scale cluster contours remain essentially unchanged.

Based on these observations, we adopt a more economical strategy: we train a single GCN on the 2D Ising model and then apply it directly to binarized $q$-state Potts configurations. Specifically, we train on the $L=128$ Ising model using configurations at $T/J\in[0.01,0.05]$ (ordered) and $T/J\in[100,104]$ (disordered), with Ising spins $s_i=\pm 1$ as node features. After training, all parameters are frozen. We then generate equilibrium configurations for planar Potts models with $q\in\{2,\dots,8, 10, 15\}$ over $T/J\in[0.5,1.5]$, binarize each configuration, and feed it into this Ising pretrained GCN to obtain $p_\text{o}(T)$ and $p_\text{d}(T)$.

\begin{figure}[t]
  \centering
  \includegraphics[width=1\linewidth]{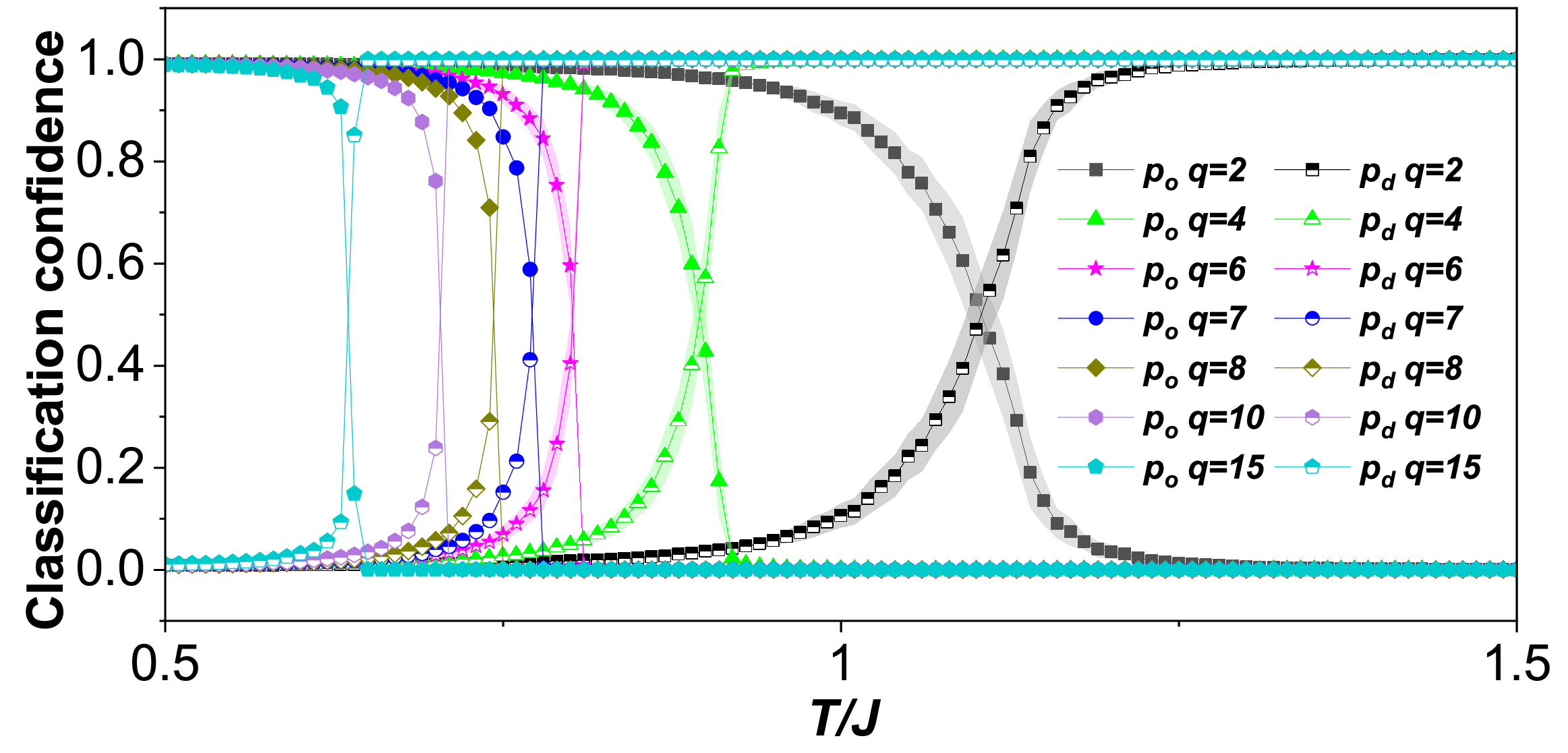}
\caption{The confidence curves $p_\text{o}(T)$ and $p_\text{d}(T)$ are obtained by applying an Ising-pretrained GCN to binarized 2D Potts configurations for $q\in\{2,\dots,8, 10, 15\}$ ($L=128$) across the temperature range $T/J\in[0.5,1.5]$.}

  \label{fig:planar_q2to6_confidence_IsingGCN_cn}
\end{figure}

Fig.~\ref{fig:planar_q2to6_confidence_IsingGCN_cn} presents the GCN confidence curves, $p_\text{o}(T)$ and $p_\text{d}(T)$, for the two-dimensional Potts model with $q\in\{2,\dots,8, 10, 15\}$. As temperature increases, $p_\text{o}(T)$ decreases monotonically while $p_\text{d}(T)$ rises accordingly. The crossing point where $p_\text{o}(T)\simeq p_\text{d}(T)$ shifts to lower temperatures as $q$ grows——a trend that aligns with the exact decrease of $T_c^{2\mathrm{D}}(q)$. Moreover, for $q > 4$, the transition region becomes noticeably sharper, reflecting the expected first-order character of the phase transition.

\textit{Bootstrap estimation of $T_c$ and uncertainties.}
To extract quantitative critical temperatures $T_c(q)$ from the confidence curves, one could fit the crossing point $p_\text{o}(T)=p_\text{d}(T)$ within a narrow window. However, for first-order transitions the confidence curves can change extremely rapidly near the crossing, making conventional error propagation unstable. We therefore employ bootstrap resampling ~\cite{efron1992bootstrap,kawano1995bootstrap,LCYThesis}: at each temperature we resample the classification outcomes with replacement to generate multiple pseudo-datasets, extract the crossing point for each, and collect the bootstrap samples $\{T_c^{(b)}(q)\}$. The mean and standard deviation of this set are then reported as the final estimate $T_c(q)$ and its statistical uncertainty, respectively. Further implementation details are provided in Appendix \ref{app:bootstrap}.

\begin{figure}[t]
  \centering
  \includegraphics[width=1\linewidth]{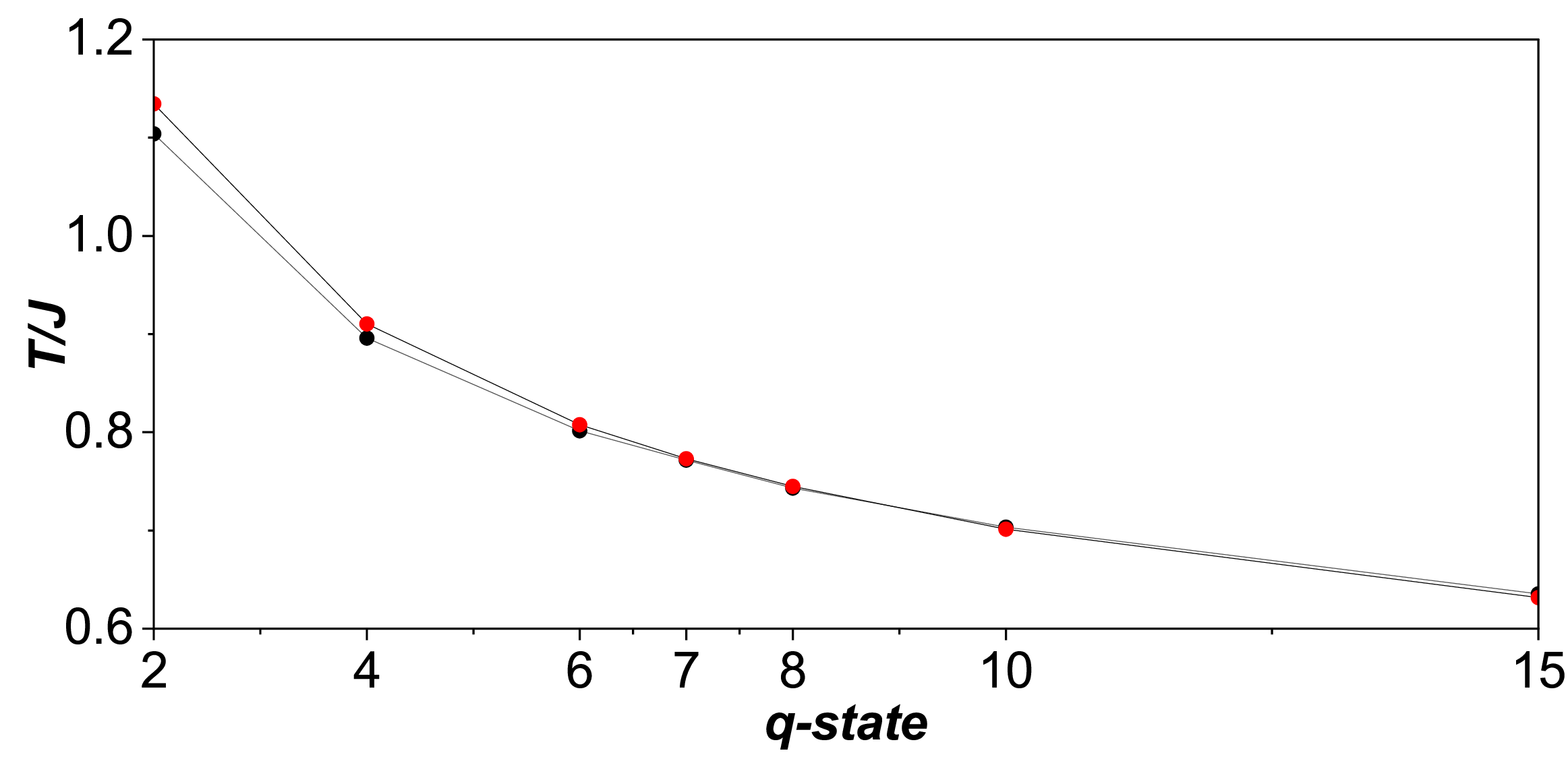}
  \caption{$T_c(q)$ for the 2D Potts model ($L=128$).
  Black: values obtained via an Ising-pretrained GCN and bootstrap; Red: exact results $T_c^{2\mathrm{D}}(q)/J=1/\ln(1+\sqrt{q})$.}
  \label{fig:Tc_vs_q_planar_cn}

\end{figure}

Using this bootstrap procedure, we obtain the planar baseline critical temperatures as
 $T_c^{(2\mathrm{D})}(q)/J=\{1.1038(8),\,0.9877(18),\,0.8958(4),\,0.8439(5),\,0.8016(1), \\
 \,0.7715(2),\,0.7429(1), \,0.7034(1),\,0.6354(1)\}$ for $q\in\{2,\dots,8, 10, 15\}$ (uncertainties refer to the last two digits). Fig.~\ref{fig:Tc_vs_q_planar_cn} compares these values with the exact results $T_c^{2\mathrm{D}}(q)/J=1/\ln(1+\sqrt{q})$, showing excellent agreement. The quoted uncertainties reflect only statistical fluctuations arising from finite sampling. The residual $\sim\!1\%$ deviation is attributed to a small systematic bias inherent to the GCN-based estimator, which remains stable against variations in network hyperparameters (see Appendix).

The planar benchmarks establish that (i) GCNs can locate critical points of regular 2D Potts models accurately, and (ii) after binarization an Ising pretrained GCN transfers across different $q$ without retraining while retaining critical information. We therefore apply the same framework to the spherical Potts model to quantify how curvature and spherical topology shift $T_c$ relative to planar baselines.

\subsection{Spherical GCN analysis and geometric effects}

\textit{Spherical transfer across $q$ via an Ising-pretrained GCN.}

On the sphere, each configuration is treated as a scalar field defined on the spherical Fibonacci lattice, where nodes correspond to lattice sites, spins serve as node features, and edges connect site pairs separated by a chord distance smaller than the cutoff $r_c$.
We first train the GCN exclusively on the spherical Ising model ($q=2$) with parameters $R=10$, $N=1000$, and $r_c/R=0.1298$. Configurations from the low-temperature regime $T/J\in[0.01,0.05]$ are labeled as ordered, while those from the high-temperature regime $T/J\in[100,104]$ are labeled as disordered. Once trained, the network parameters are frozen. This frozen GCN is then applied directly to binarized spherical Potts configurations for $q>2$.

\begin{figure}[htbp]
\centering
\includegraphics[width=1.5in]{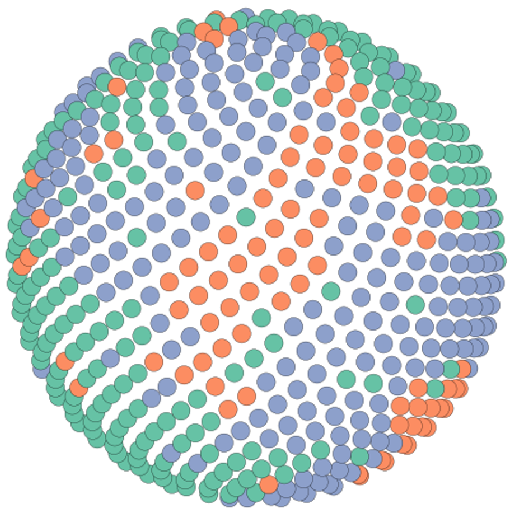}
\hspace{0.2in}
\includegraphics[width=1.5in]{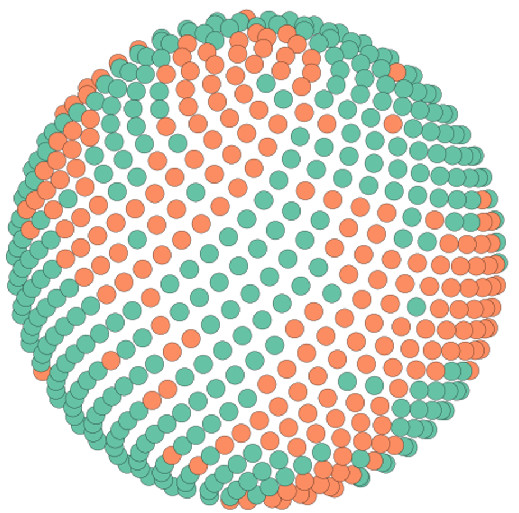}\\
\caption{Spherical Potts configurations and their binarized versions ($q=3$; original left, binarized right) at $T/J=1.0$.}
\label{fig:sphere_binary_q3}
\end{figure}

Typical spherical Potts configurations for $q=3$ at $T/J=1.0$ and their binarized counterparts are displayed in Fig.~\ref{fig:sphere_binary_q3}. The large-scale cluster boundaries and characteristic length scales are preserved after binarization, whereas the original Potts states are compressed into two effective categories. This visual consistency under binarization provides a geometric justification for transferring the spherical Ising-pretrained GCN to multi-state Potts models on the sphere.

\begin{figure}[ht]
\centering
  \includegraphics[width=3.5in]{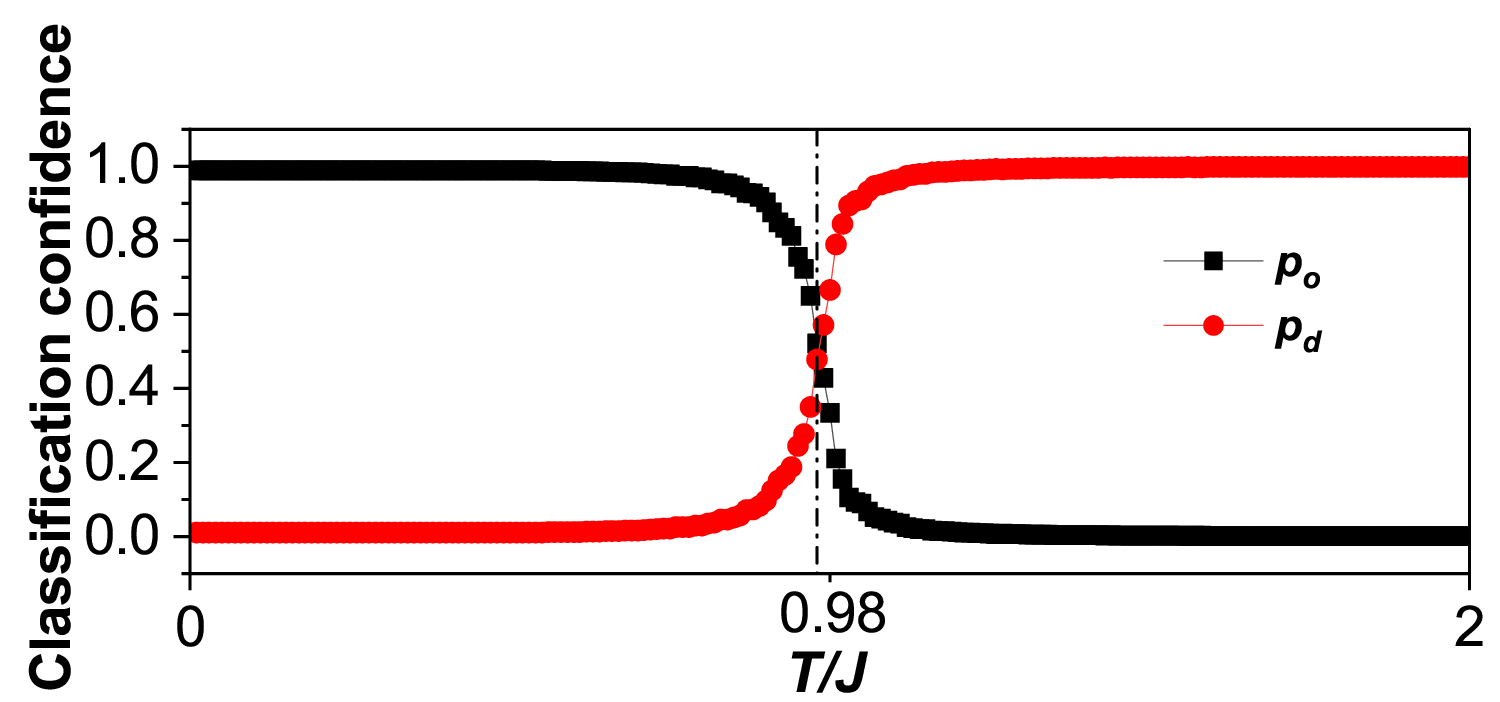}\\
  \includegraphics[width=3.5in]{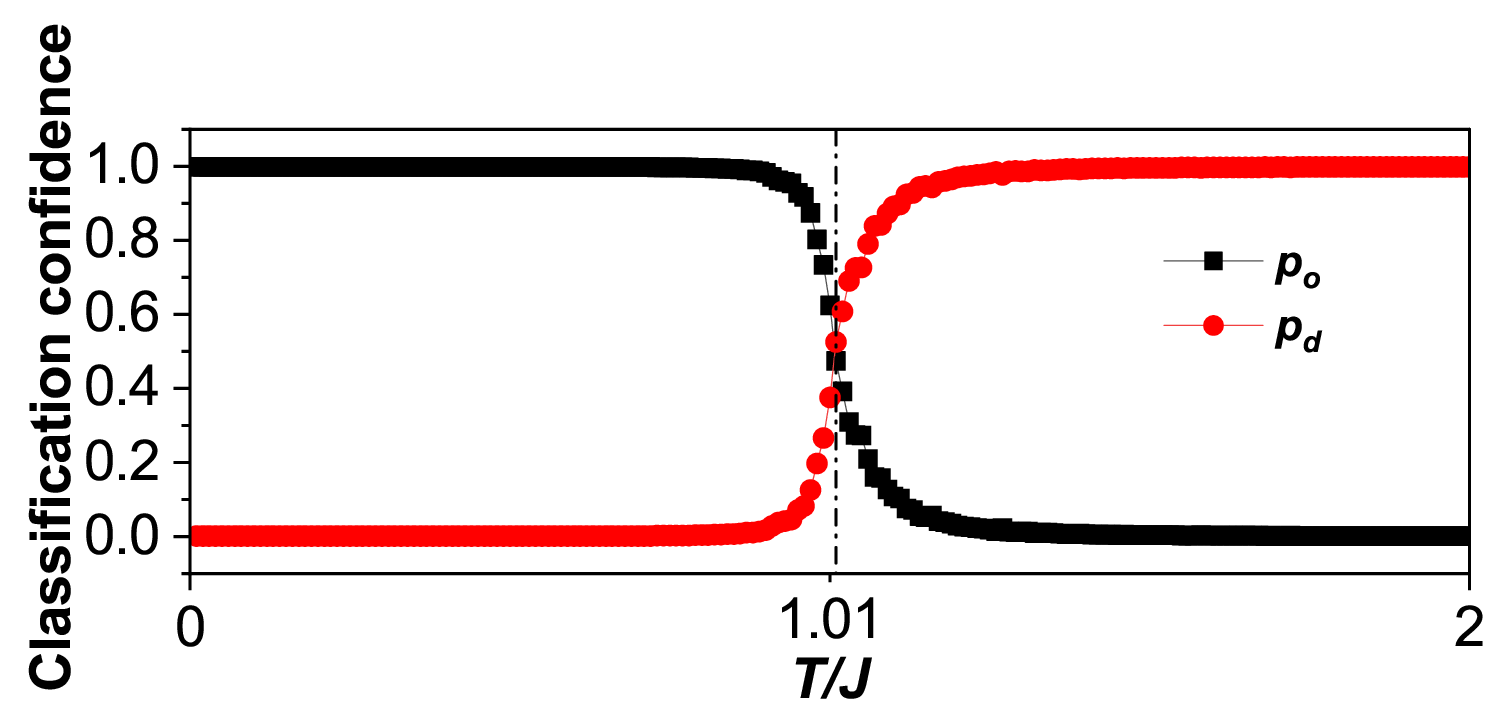}
\caption{Spherical $q=3$ Potts GCN confidence curves: direct training (top) vs transfer from spherical-Ising pretraining on binarized data (bottom).}
\label{fig:sphere_q3_GCN_compare}
\end{figure}

\textit{Benchmark of transfer at $q=3$.}
To validate the transfer strategy on the spherical graph, we first perform a benchmark at $q=3$. The upper panel of Fig.~\ref{fig:sphere_q3_GCN_compare} displays confidence curves obtained from a GCN trained directly on spherical $q=3$ configurations, which gives a crossing at $T_c/J \approx 0.98$. The lower panel illustrates our primary transfer approach: spherical $q=3$ configurations are first binarized and then fed into the frozen GCN that was pretrained on the spherical Ising model. This procedure yields a crossing at $T_c/J \approx 1.01$. The two estimates agree within a few percent, confirming the validity of the transfer strategy based on binarization.

\begin{figure}[ht]
\centering
  \includegraphics[width=3.5in]{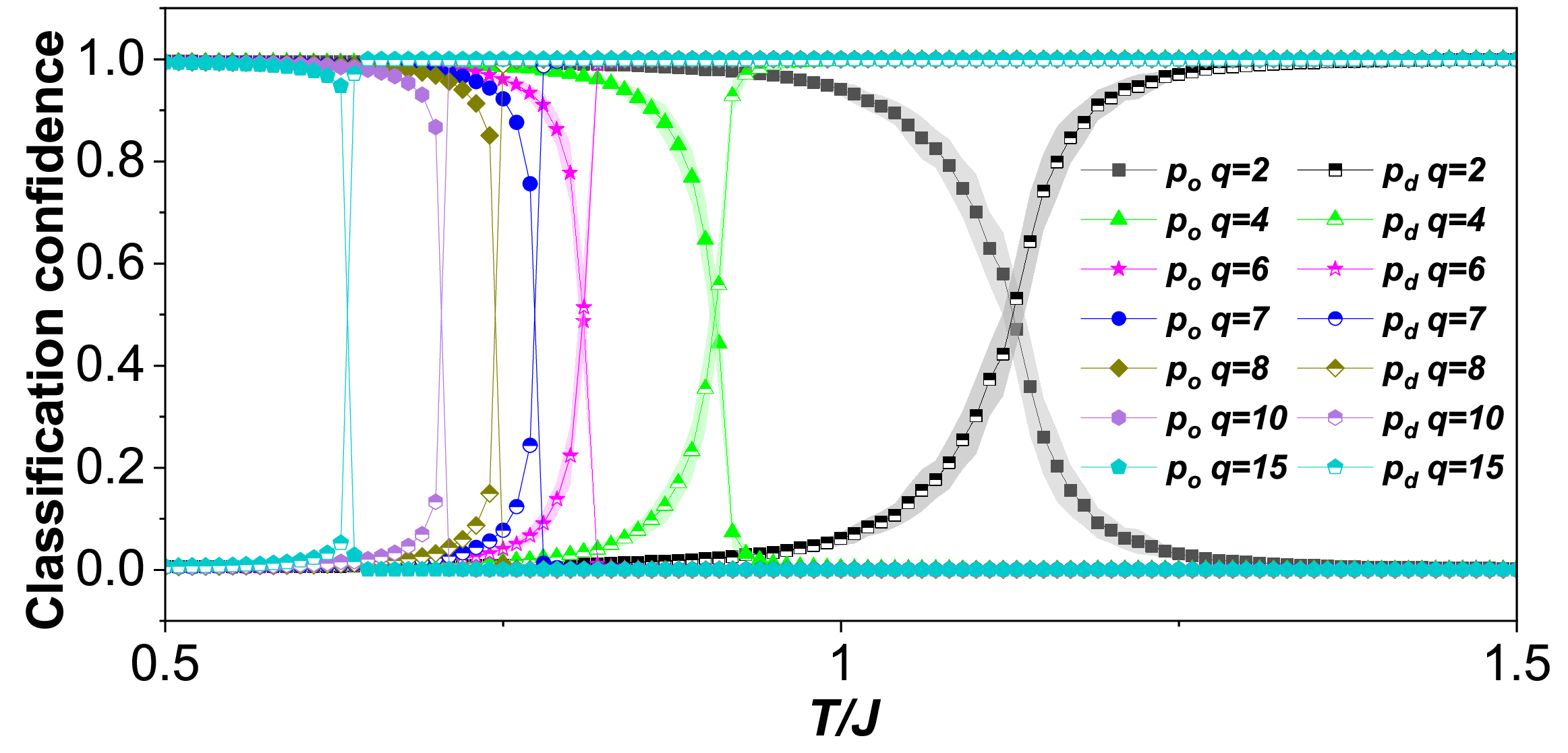}\\
  \includegraphics[width=3.5in]{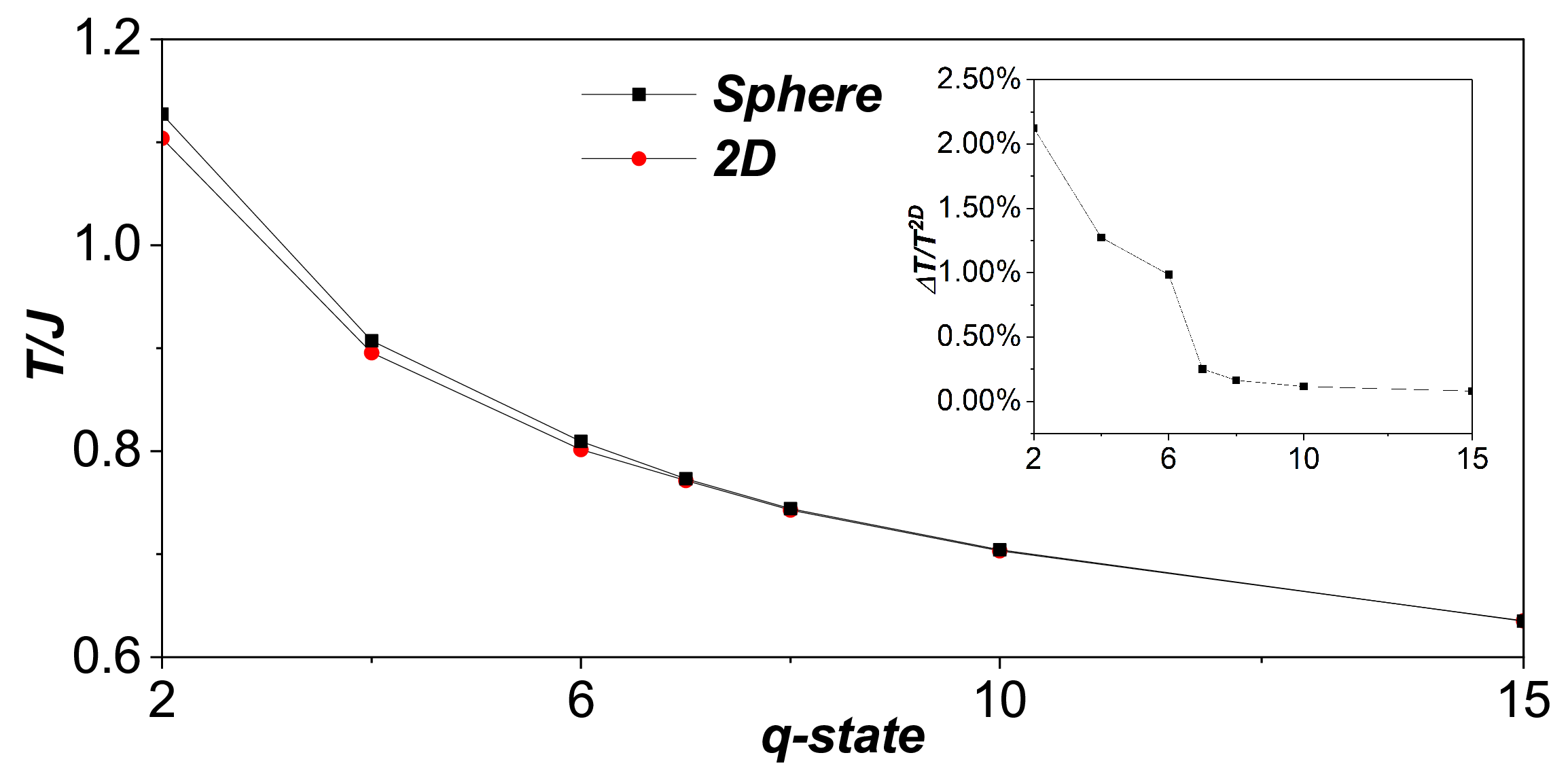}
\caption{Spherical Potts results compared with planar baselines.
Top: GCN confidence curves $p_\text{o}(T)$ and $p_\text{d}(T)$ for binarized $q\in\{2,\dots,8, 10, 15\}$ on the
fixed spherical Fibonacci graph ($N=128\times 128=16384$, $R=10$, and $r_c/R=0.0321$).
Bottom: critical temperatures versus $q$: spherical $T_c^{\mathrm{sph}}(q)$
(black) compared with planar $T_c^{2\mathrm{D}}(q)$ from the $L=128$ square
lattice (red). The inset shows the relative difference between the spherical and planar critical temperatures $\Delta T$, normalized by the planar critical temperature $T_c^{2\mathrm{D}}$.}

\label{fig:sphere_confidence_N1000}
\end{figure}

After completing the above validation, we apply the same spherical-Ising-pretrained GCN to the spherical Potts model with $q\in\{2,\dots,8, 10, 15\}$ on a fixed graph. To enable a direct comparison with the planar results for 
$T_c(q)$ of the two-dimensional Potts model at $L=128$, the geometry is fixed as follows: $N=128\times 128=16384$, $R=10$, and $r_c/R=0.0321$. The coordination statistics for this geometry are $(N_3,N_4,N_5)=(210,15658,516)$, yielding an average coordination $\bar z\simeq 4.018$. Thus approximately $95.6\%$ of nodes are four-coordinated, while the remaining $\sim 4.4\%$ are three- or five-coordinated defects
Fig.~\ref{fig:sphere_confidence_N1000} (top) shows the resulting GCN confidence curves $p_\text{o}(T)$ and $p_\text{d}(T)$. The crossing temperature $p_\text{o}(T)\simeq p_\text{d}(T)$ shifts to lower values as $q$ increases, and the crossover becomes markedly sharper for $q>4$, consistent with a first-order transition.  Bootstrap analysis gives $T_c^{\mathrm{sph}}(q)/J = \{1.1273(6),\,0.9976(10),\,0.9072(3),\,0.8519(3),\\ \,0.8095(3),\,0.7734(1),\,0.7441(1),\,0.7042(1),\,0.6349(1)\}$ for $q\in\{2,\dots,8, 10, 15\}$. For $q=2$, the extracted $T_c$ is consistent with our earlier spherical Fibonacci-lattice Ising study ~\cite{Zhou_2025}.

In contrast to the square lattice, where finite-size effects are often substantial, our earlier spherical Ising study indicated that increasing the number of Fibonacci sites $N$ on a fixed sphere produces only a weak change in the inferred transition temperature under the same graph-construction protocol ~\cite{Zhou_2025}. Based on this, we forgo an explicit finite-size analysis for the spherical Potts models here. Instead, we adopt $N=16384$ as a representative spherical discretization and use it to quantify the geometric shift relative to planar benchmarks.

Comparing the spherical estimates at $N=16384$ with the planar baselines,
$T_c^{(2\mathrm{D})}(q)/J=\{1.1038,\,0.9877,\,0.8958,\,0.8439,\,0.8016,\,0.7715,\,0.7429, \\ \,0.7034,\,0.6354\}$ for $q\in\{2,\dots,8, 10, 15\}$,
we obtain relative shifts
$(T_c^{\mathrm{sph}}-T_c^{(2\mathrm{D})})/T_c^{(2\mathrm{D})}=\Delta T /T_c^{2\mathrm{D}}\approx\{2.13\%,\,1.00\%,\,1.27\%,\,0.95\%,\,0.98\%,\,0.25\%,\,0.16\%, \\ \,0.11\%,\,-0.08\%\}$
for the same sequence of $q$. The spherical-planar difference therefore decreases rapidly with $q$: For 
$q<6$, the difference is approximately $1\%$, whereas for $q \geq 7$ it drops sharply to below $0.25\%$. 

This behavior can be naturally understood in terms of the weakly first-order transition mechanism of the two-dimensional Potts model for $q>4$. It is well established that $q=4$ marks the critical boundary between continuous and first-order phase transitions. For $q>4$, the transition is first order in the strict sense; however, when $q$ is only slightly larger than $4$ (typically $q=5,6$), the first-order character is extremely weak and accompanied by pronounced pseudocritical behavior. In this regime, the latent heat is very small and the correlation length near the transition becomes anomalously large, so that observables such as the specific heat and susceptibility exhibit an apparent, near-divergent behavior characteristic of a continuous transition on finite length scales ~\cite{binder1981static, EBuddenoir1993, iino2019detecting}. From a renormalization group perspective, this weakly first-order behavior originates from the annihilation of fixed points and the associated walking dynamics, which generate a broad quasi-critical (crossover) regime at finite scales, while the system ultimately flows to a genuine first-order transition at sufficiently large scales ~\cite{10.21468/SciPostPhys.5.5.050}.

Within this broad quasi-critical regime, the weak geometric perturbations introduced by the spherical Fibonacci graph, global curvature and a small fraction of coordination defects, can more readily modify fluctuations and cluster geometry. As a result, the apparent critical temperature $T_c$ extracted from the classifier crossing is shifted at the percent level. As $q$ increases, the first-order character strengthens, and the correlation length decreases rapidly. The associated crossover window thus becomes significantly narrower, effectively reducing the sensitivity of the extracted $T_c$ to geometric perturbations. This provides a natural explanation for why the spherical-planar difference drops rapidly to the $10^{-3}$ level for $q\geq7$.

\textit{Effect of cutoff and average coordination at fixed $N$.}
Finally, we investigate how local connectivity influences $T_c$ by varying the cutoff $r_c$ while keeping $R=10$ and $N=1000$ fixed.
The resulting coordination distributions are listed in Table~\ref{tab:coord_rc} for three choices $r_c/R=0.1298,\,0.1500,\,0.1700$.
For $r_c/R=0.1298$ the graph is predominantly four-neighbor ($\bar z\simeq 3.998$);
for $r_c/R=0.1500$ five- and six-neighbor sites dominate ($\bar z\simeq 5.71$);
and for $r_c/R=0.1700$ six-, seven-, and eight-neighbor sites dominate ($\bar z\simeq 6.31$).
Thus, by tuning $r_c$ we interpolate between a sparse, nearly four-neighbor network and a densely connected graph on the same spherical surface.

\begin{table}[t]
  \centering
\caption{Coordination statistics for spherical Fibonacci graphs ($R=10$, $N=1000$) at different cutoffs $r_c/R$.}
  \setlength{\tabcolsep}{4pt}
  \renewcommand{\arraystretch}{1.1}
  \begin{tabular}{|c|c|c|c|c|c|c|}
  \hline
   \textbf{$r_c/R$} & \makecell{$N_3$\\ ($w_3$)} & \makecell{$N_4$\\ ($w_4$)} & \makecell{$N_5$\\ ($w_5$)} & \makecell{$N_6$\\ ($w_6$)} & \makecell{$N_7$\\ ($w_7$)} & \makecell{$N_8$\\ ($w_8$)} \\
  \hline
  0.1298 & \makecell{76\\ (7.6\%)} & \makecell{850\\ (85\%)} & \makecell{74\\ (7.4\%)} & -- & -- & -- \\
  \hline
  0.1500 & -- & \makecell{26\\ (2.6\%)} & \makecell{240\\ (24\%)} & \makecell{734\\ (73.4\%)} & -- & -- \\
  \hline
  0.1700 & -- & -- & \makecell{8\\ (0.8\%)} & \makecell{722\\ (72.2\%)} & \makecell{224\\ (22.4\%)} & \makecell{46\\ (4.6\%)} \\
  \hline
  \end{tabular}
  \label{tab:coord_rc}
\end{table}

The corresponding $T_c^{\mathrm{sph}}(q;r_c)$ values are listed in Table~\ref{tab:Tc_vs_q_r}. For every $q$, $T_c^{\mathrm{sph}}$ rises significantly with $r_c$. Taking $q=3$ as an example, $T_c^{\mathrm{sph}}(3;r_c/R=0.1298,0.1500,0.1700)/J\simeq(1.0085,\,1.5254,\,1.7058)$ an increase by roughly a factor of two as the connectivity changes from sparse to dense. This trend holds for all $q$: as the average coordination $\bar z$ grows from $\sim4.0$ to $\sim6.3$, $T_c^{\mathrm{sph}}$ is systematically shifted upward, while the monotonic decrease of $T_c$ with $q$ is preserved. These findings indicate that variations in local connectivity predominantly rescale the overall magnitude of $T_c$, whereas the spin multiplicity $q$ governs the transition order and the relative $q$-dependence of the critical temperature.

\begin{figure}[ht]
\centering
  \includegraphics[width=3.5in]{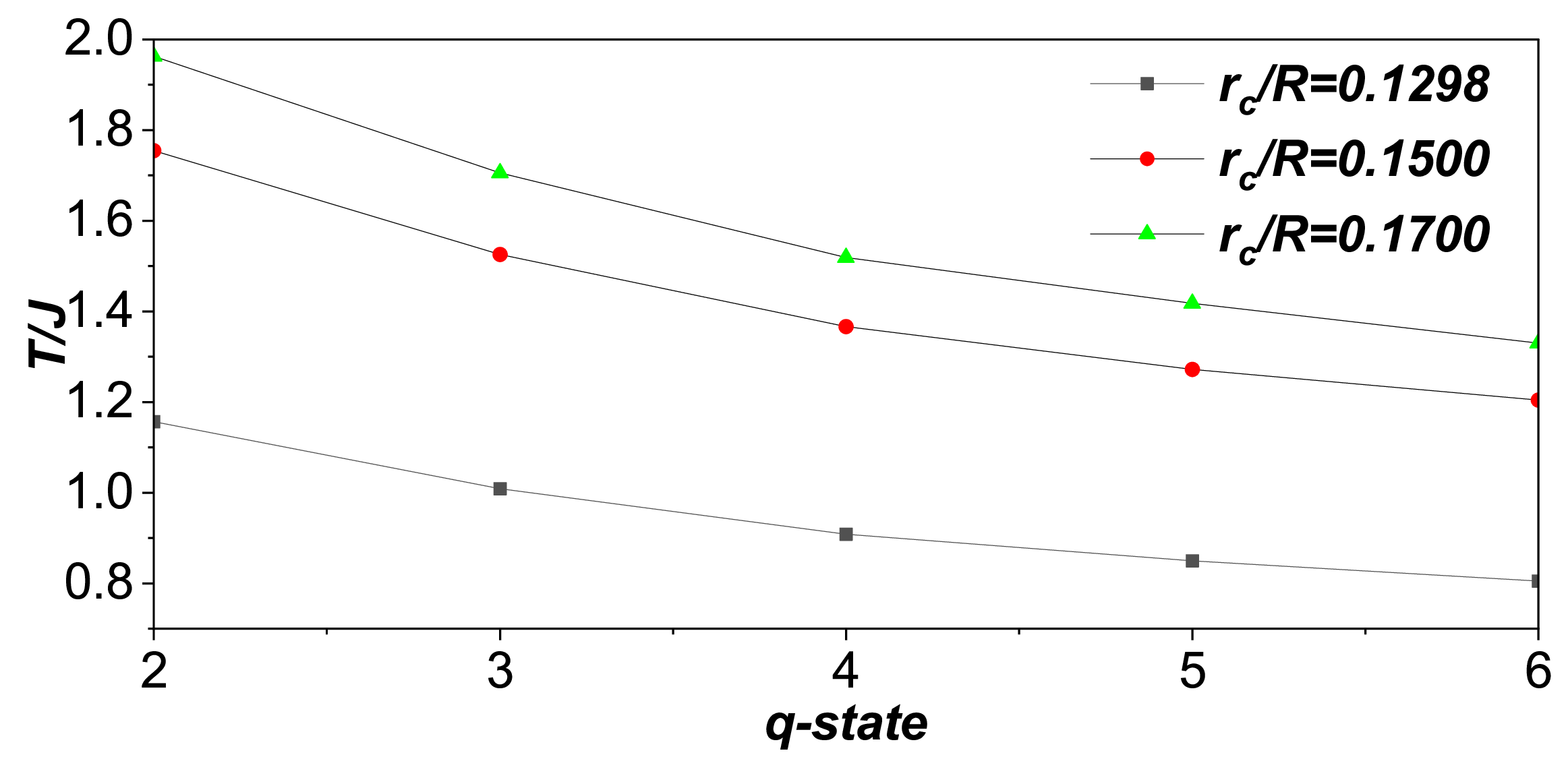}
\caption{Critical temperature $T_c^{\mathrm{sph}}(q;r_c)$ versus $q$ for different cutoffs $r_c/R=0.1298,0.1500,0.1700$ at fixed $R=10$ and $N=1000$.}
\label{fig:sphere_Tc_vs_q_rc}
\end{figure}

\begin{table}[t]
  \centering
\caption{Spherical Potts critical temperatures $T_c^{\mathrm{sph}}(q;r_c)$ at $N=1000$ for different $r_c/R$.}

  \setlength{\tabcolsep}{4pt}
  \renewcommand{\arraystretch}{1.1}
  \begin{tabular}{cccc}
    \hline\hline
    $q$ & $r_c/R=0.1298$ & $r_c/R=0.1500$ & $r_c/R=0.1700$ \\
    \hline
    2 & 1.1564(16) & 1.7544(33) & 1.9630(47) \\
    3 & 1.0085(15) & 1.5254(33) & 1.7058(32) \\
    4 & 0.9081(8)  & 1.3665(15) & 1.5191(20) \\
    5 & 0.8496(8)  & 1.2722(26) & 1.4178(11) \\
    6 & 0.8048(2)  & 1.2042(4)  & 1.3296(12) \\
    \hline\hline
  \end{tabular}
  \label{tab:Tc_vs_q_r}
\end{table}

In summary, these results demonstrate that a spherical-Ising-pretrained GCN reliably transfers to binarized spherical Potts models, producing critical temperatures that align closely with planar benchmarks. For locally four-neighbor graphs, the critical temperatures on the spherical Fibonacci lattice already show close agreement with the planar benchmarks, with curvature and coordination defects introducing only percent-level shifts. The rapid reduction of the spherical–planar difference with increasing $q$ can be naturally understood in terms of the weakly first-order nature of the Potts transition for $q>4$: when 
$q$ is only slightly above 4, the broad quasi-critical regime enhances sensitivity to weak geometric perturbations, whereas at larger $q$ the strengthened first-order character significantly narrows the crossover window, effectively reducing the sensitivity of the extracted $T_c$	to geometric effects.
At fixed $N$, varying $r_c$ primarily rescales the overall transition temperature for every $q$, leaving the functional form of $T_c(q)$ largely unchanged. Together, these observations support a clear separation between geometric influences, curvature and connectivity, and the intrinsic role of spin multiplicity $q$, all captured within a unified GCN-based framework.

\section{Conclusion}\label{V}

We have studied the ferromagnetic $q$-state Potts model on spherical Fibonacci graphs, quantifying how curvature and irregular coordination modify phase transitions relative to planar benchmarks. The spherical graphs were constructed using a chord-distance cutoff that maximizes local four-coordination, resulting in predominantly four-neighbor networks with a controlled fraction of coordination defects. Equilibrium configurations were generated via the Swendsen-Wang cluster algorithm, with Fortuin-Kasteleyn bonds recorded to visualize connectivity.

Critical temperatures were identified using graph convolutional networks (GCNs) that operate directly on the graph adjacency matrix and node spins. After validating the method on planar lattices, we introduced a transfer strategy based on spin binarization. This allows an Ising-pretrained GCN to be applied across different $q$ without retraining. Bootstrap resampling provided robust $T_c$ estimates even near first-order transitions.

For nearly four-neighbor spherical graphs, the curvature-induced shift in $T_c$ is most noticeable at small $q$ (about $1-2\%$ for $q<7$) and decreases to about $0.2\%$ or less for $q\geq 7$. consistent with the weakly first-order character of the Potts transition for $q>4$, which implies an enhanced geometric sensitivity at smaller 
$q$ and a rapid suppression of such effects as $q$ increases. At fixed $N$, raising the cutoff $r_c$ (and thus the average coordination $\bar z$) primarily shifts $T_c$ upward for all $q$, while the monotonic $q$-dependence and the order of the transition remain unchanged. These results underscore a clear separation between geometric effects, namely curvature and connectivity, and the role of spin multiplicity $q$.

More broadly, this work demonstrates that GCNs provide a unified framework for phase classification on both regular and curved graphs, using the same input representation of adjacency and spins. The combination of spherical Fibonacci discretization, cluster Monte Carlo sampling, and binarization-based transfer learning offers a scalable approach for studying multi-state models on non-Euclidean geometries. This framework can be extended to other curved manifolds, alternative graph constructions, or higher-$q$ regimes where first-order behavior becomes more pronounced. Code and graph generators are available upon reasonable request.

\section{Acknowledgments}
H. G. was supported by the Innovation Program for Quantum Science and Technology-National
Science and Technology Major Project (Grant No. 2021ZD0301904) and the National Natural Science
Foundation of China (Grant No. 12447216).  X. Y. H. was supported by the the National Natural Science
Foundation of China (Grant No. 12405008).

Zheng Zhou and Xu-Yang Hou contributed equally to this work.

\appendix

\section{Swendsen-Wang cluster update for the spherical Potts model}
\label{app1}

We generate equilibrium configurations for the ferromagnetic $q$-state Potts model on spherical Fibonacci graphs using the Swendsen--Wang (SW) cluster Monte Carlo algorithm. This section outlines one SW sweep on an irregular interaction graph.

\textit{Interaction Graph.}
The system is defined on a graph $G=(V,E)$ embedded on a sphere of radius $R$, where $|V|=N$ sites are placed at Fibonacci coordinates $\{\boldsymbol r_i\}_{i=1}^N$.
Two sites $i$ and $j$ are connected by an edge if their chord distance
\begin{equation}
d_{ij}\equiv |\boldsymbol r_i-\boldsymbol r_j|
\end{equation}
satisfies $d_{ij}<r_c$. We denote interacting pairs by $\langle i,j\rangle\in E$.
The Potts Hamiltonian on this graph is
\begin{equation}
H(\{\sigma\})=-J\sum_{\langle i,j\rangle}\delta_{\sigma_i,\sigma_j},
\qquad \sigma_i\in\{0,1,\dots,q-1\},\quad J>0,
\end{equation}
with inverse temperature $\beta\equiv 1/T$ (setting $k_B=1$).

\textit{Fortuin--Kasteleyn Bond Activation.}
The SW algorithm employs auxiliary bond variables in the Fortuin--Kasteleyn (FK) representation.
Given a spin configuration $\{\sigma_i\}$, a bond is activated on edge $\langle i,j\rangle$ only if the two spins are equal:
\begin{equation}
b_{ij}=
\begin{cases}
1, & \text{with probability } p_{\mathrm{add}}=1-e^{-\beta J}, \text{ if } \sigma_i=\sigma_j,\\[3pt]
0, & \text{if } \sigma_i\neq \sigma_j.
\end{cases}
\label{eq:padd}
\end{equation}
Thus, bonds appear exclusively within regions of like spins.

\textit{Cluster Identification and Recoloring (One SW Sweep).}
Let $E_{\mathrm{act}}=\{\langle i,j\rangle\in E:\, b_{ij}=1\}$ be the set of activated bonds.
The subgraph $(V,E_{\mathrm{act}})$ decomposes into connected components (FK clusters) $\{C_\alpha\}$.

A single SW sweep proceeds as follows:
\begin{enumerate}
    \item \textbf{Bond Construction:} Sample all $b_{ij}$ according to Eq.~\eqref{eq:padd}.
    \item \textbf{Cluster Recoloring:} For each FK cluster $C_\alpha$, choose a new Potts state $\sigma'_\alpha\in\{0,1,\dots,q-1\}$ uniformly at random and assign
    \begin{equation}
    \sigma_i \leftarrow \sigma'_\alpha,\qquad \forall\, i\in C_\alpha.
    \end{equation}
\end{enumerate}
By reassigning entire clusters collectively, the update performs nonlocal moves that efficiently decorrelate long-wavelength fluctuations near criticality.

\textit{Detailed Balance and Efficiency.}
The SW update can be viewed as alternating conditional sampling in the joint FK spin--bond ensemble: sample bonds conditioned on spins, then sample spins conditioned on bonds. This construction ensures the Markov chain satisfies detailed balance and has the correct Boltzmann distribution as its stationary distribution. Compared to local single-spin updates, cluster flips substantially mitigate critical slowing down by updating strongly correlated regions as a unit.

\section{ GCN architecture and training details}\label{app2}

We treat each spin configuration as a graph signal defined on the interaction graph. The graph convolutional network (GCN) takes the graph connectivity and node spins as input and outputs two confidence scores for the ordered and disordered phases.

\textit{Interaction Graph and Matrices.}
For a spherical Fibonacci point set $\{\boldsymbol r_i\}_{i=1}^N$ on a sphere of radius $R$, we construct an unweighted, undirected interaction graph $\mathcal G=(\mathcal V,\mathcal E)$ with $\mathcal V=\{1,\dots,N\}$. Two vertices $i$ and $j$ are connected if their chord distance $d_{ij}=|\boldsymbol r_i-\boldsymbol r_j|$ is smaller than a cutoff radius $r_c$. The adjacency matrix $\boldsymbol A\in\{0,1\}^{N\times N}$ is defined by
\begin{equation}
A_{ij}=
\begin{cases}
1, & i\neq j\ \text{and}\ d_{ij}<r_c,\\
0, & \text{otherwise},
\end{cases}
\end{equation}
and the degree matrix is $\boldsymbol D=\mathrm{diag}(d_1,\dots,d_N)$ with $d_i=\sum_j A_{ij}$.
We use the random-walk normalized Laplacian
\begin{equation}
\boldsymbol L^{\mathrm{rw}}=\boldsymbol D^{-1}(\boldsymbol D-\boldsymbol A).
\end{equation}
Only $\boldsymbol A$ (hence $\boldsymbol D$ and $\boldsymbol L^{\mathrm{rw}}$) is used by the GCN; the explicit coordinates $\{\boldsymbol r_i\}$ are used only to construct the connectivity.

\textit{Node Features.}
For each Monte Carlo configuration, we form a node-feature matrix $\boldsymbol X\in\mathbb R^{N\times F}$. For the Ising model, $F=1$ and $X_i=s_i\in\{-1,+1\}$. For the Potts model, we either use the raw Potts labels $X_i=\sigma_i$ (for direct training at a fixed $q$) or the binarized variable $X_i=s_i$ defined below (for transfer learning across $q$).

\textit{Graph Convolution Layer.}
A single graph-convolution layer is implemented as
\begin{equation}\label{eq:gcn_layer}
\boldsymbol H=\mathrm{ReLU}\!\left(\boldsymbol L^{\mathrm{rw}}\boldsymbol X\,\boldsymbol W_h+\boldsymbol b_h\right),
\end{equation}
where $\boldsymbol W_h\in\mathbb R^{F\times F_h}$ is the trainable weight matrix, $\boldsymbol b_h\in\mathbb R^{F_h}$ is a trainable bias (broadcast to all nodes), $\mathrm{ReLU}(x)=\max(x,0)$ is applied elementwise, and $\boldsymbol H\in\mathbb R^{N\times F_h}$ is the hidden node embedding.

\textit{Graph-Level Readout and Confidence Output.}
To classify an entire configuration, we aggregate node embeddings by global average pooling,
\begin{equation}
\boldsymbol h_{\mathcal G}=\frac{1}{N}\sum_{i=1}^N \boldsymbol H_i \in\mathbb R^{F_h},
\end{equation}
followed by a fully connected layer producing two logits,
\begin{equation}
\boldsymbol z=\boldsymbol W_\text{o}\,\boldsymbol h_{\mathcal G}+\boldsymbol b_\text{o}\in\mathbb R^{2},
\end{equation}
and a softmax output
\begin{equation}
(p_\text{o},p_\text{d})=\mathrm{softmax}(\boldsymbol z),
\qquad
\mathrm{softmax}(z_k)=\frac{\me^{z_k}}{\sum_{\ell=1}^2 \me^{z_\ell}}.
\end{equation}
Here $p_\text{o}$ and $p_\text{d}$ are interpreted as the confidences for the ordered and disordered phases, respectively.

\textit{Supervised Training at Extreme Temperatures.}
The network is trained in a supervised manner using configurations generated at extreme temperatures: low-$T$ configurations are labeled as ordered $(1,0)$, and high-$T$ configurations as disordered $(0,1)$. Denoting the ground-truth label by $\boldsymbol y\in\{(1,0),(0,1)\}$, we minimize the cross-entropy loss
\begin{equation}
\mathcal L=-\sum_{k\in\{o,d\}} y_k\ln p_k.
\end{equation}
After training converges, the network parameters are frozen. Inference is then performed on configurations sampled across the target temperature range, producing the confidence curves $p_\text{o}(T)$ and $p_\text{d}(T)$.

\textit{Binarization (Potts $\to$ Ising) for Transfer Across $q$.}
To use a single Ising-pretrained classifier for different $q$, we map the Potts spin $\sigma_i\in\{0,1,\dots,q-1\}$ to a binary variable $s_i=\pm1$ via
\begin{equation}\label{eq:binarize_app}
s_i=
\begin{cases}
-1, & \sigma_i \in \{0,1,\dots,\lfloor (q-1)/2\rfloor\},\\
+1, & \sigma_i \in \{\lfloor (q-1)/2\rfloor+1,\dots,q-1\}.
\end{cases}
\end{equation}
This rule is unambiguous for both even and odd $q$. In the transfer setting, we apply Eq.~\eqref{eq:binarize_app} to each Potts configuration to obtain $\{s_i\}$, and feed $\boldsymbol X=(s_1,\dots,s_N)^{T}$ into the frozen Ising-pretrained GCN on the same interaction graph.

\textit{Estimating $T_c$ from Confidence Curves.}
For each $q$, we identify the transition region as the temperature where $p_\text{o}(T)\simeq p_\text{d}(T)$. On a discrete temperature grid, this corresponds to the interval where $p_\text{o}(T)-p_\text{d}(T)$ changes sign. The critical temperature $T_c$ is then extracted by interpolation or the bootstrap procedure described in the main text.

\section{Bootstrap estimation of $T_c$ and systematic bias discussion}\label{app:bootstrap}

For each temperature $T_i$, the trained GCN outputs a two-component logit vector $\boldsymbol{o}=(o_\text{o},o_\text{d})$ per configuration. We convert logits to classification confidences via the softmax function:
\begin{equation}
p_\text{o}=\frac{\me^{o_\text{o}}}{\me^{o_\text{o}}+\me^{o_\text{d}}},\qquad
p_\text{d}=\frac{\me^{o_\text{d}}}{\me^{o_\text{o}}+\me^{o_\text{d}}},\qquad (p_\text{o}+p_\text{d}=1).
\end{equation}
The curves $p_\text{o}(T)$ and $p_\text{d}(T)$ represent the ordered- and disordered-phase confidences, respectively. Since $p_\text{o}+p_\text{d}=1$, the criterion $p_\text{o}=p_\text{d}$ is equivalent to $p_\text{o}=0.5$.

\textit{Temperature Grouping and Sample Mean.}
Data are grouped by temperature: for each $T_i$, we have $N_{\mathrm{samp}}$ equilibrium configurations ($N_{\mathrm{samp}}=100$ in our implementation). Let $\{p_{o,i}^{(k)}\}_{k=1}^{N_{\mathrm{samp}}}$ denote the ordered-confidence values predicted at $T_i$. The empirical mean confidence is
\begin{equation}
\widehat{p}_\text{o}(T_i)=\frac{1}{N_{\mathrm{samp}}}\sum_{k=1}^{N_{\mathrm{samp}}} p_{o,i}^{(k)} .
\end{equation}

\textit{Bootstrap Resampling.}
To estimate the statistical uncertainty of $T_c$ arising from finite sampling at each temperature, we apply nonparametric bootstrap independently at every $T_i$.
For the $b$-th bootstrap replica, we resample with replacement:
\begin{equation}
\{p_{o,i}^{(k)}\}_{k=1}^{N_{\mathrm{samp}}}\ \longrightarrow\
\{p_{o,i}^{*(k)}\}_{k=1}^{N_{\mathrm{samp}}},
\end{equation}
and compute the resampled mean:
\begin{equation}
\widehat{p}_\text{o}^{(b)}(T_i)=\frac{1}{N_{\mathrm{samp}}}\sum_{k=1}^{N_{\mathrm{samp}}} p_{o,i}^{*(k)} .
\end{equation}
Repeating for all $i$ yields one bootstrap realization of the confidence curve $\widehat{p}_\text{o}^{(b)}(T)$.

\textit{Extracting $T_c$ from Each Bootstrap Curve.}
For each bootstrap curve, we locate the transition as the first temperature interval where $\widehat{p}_\text{o}^{(b)}(T)$ crosses $0.5$. Define
\begin{equation}
\Delta_i^{(b)}=\widehat{p}_\text{o}^{(b)}(T_i)-0.5.
\end{equation}
A crossing occurs between $T_i$ and $T_{i+1}$ if $\Delta_i^{(b)}\Delta_{i+1}^{(b)}<0$. If multiple crossings appear due to noise, we take the lowest-temperature one (the first sign change) to define $T_c^{(b)}$.

Given a crossing bracket $(T_i,T_{i+1})$, we estimate $T_c^{(b)}$ by linear interpolation:
\begin{equation}
T_c^{(b)}
= T_i+\frac{0.5-\widehat{p}_\text{o}^{(b)}(T_i)}{\widehat{p}_\text{o}^{(b)}(T_{i+1})-\widehat{p}_\text{o}^{(b)}(T_i)}
\,(T_{i+1}-T_i).
\label{eq:boot_Tc_interp}
\end{equation}
Bootstrap replicas showing no crossing over the scanned temperature window are discarded.

\textit{Final Estimate and Uncertainty.}
Collecting all valid bootstrap estimates $\{T_c^{(b)}\}_{b=1}^{N_{\mathrm{boot}}}$ (with $N_{\mathrm{boot}}=2000$), we report
\begin{equation}
\overline{T_c}=\frac{1}{N_{\mathrm{eff}}}\sum_{b=1}^{N_{\mathrm{eff}}}T_c^{(b)},\quad
\delta T_c=\sqrt{\frac{1}{N_{\mathrm{eff}}-1}\sum_{b=1}^{N_{\mathrm{eff}}}\left(T_c^{(b)}-\overline{T_c}\right)^2},
\end{equation}
where $N_{\mathrm{eff}}$ is the number of bootstrap replicas exhibiting a crossing.
The uncertainty $\delta T_c$ quantifies statistical fluctuations due to finite configuration sampling at each temperature; it does not include systematic effects from the finite temperature step $\Delta T$, finite-size corrections, or model/architecture dependence.
\\
\\
\\
\\
\\
\\
\\
\\
\\
\\
\\
\bibliography{Review,Review1,review_new,Review_spotts}
\end{document}